\documentclass[%
 reprint,
 amsmath,amssymb,
 aps,
 prl,
]{revtex4-2}

\usepackage{graphicx}
\usepackage{dcolumn}
\usepackage{bm}

\begin{document}

\preprint{APS/123-QED}

\title{The impact of homophily on digital proximity tracing}

\author{Giulio Burgio$^1$}

\author{Benjamin Steinegger$^1$}%

\author{Giacomo Rapisardi$^{1,2}$}%
\author{Alex Arenas$^1$}%
\email{alex.arenas@urv.cat}
\affiliation{%
 $^1$Departament d’Enginyeria Informàtica i Matemàtiques, Universitat Rovira i Virgili, 43007 Tarragona, Spain \\
 $^2$Barcelona Supercomputing Center (BSC)}%

\date{\today}

\begin{abstract}
We study how homophily of human physical interactions affects the efficacy of digital proximity tracing. Analytical results show a non monotonous dependence of the reproduction number with respect to the mixing rate between individuals that adopt the contact tracing app and the ones that do not. Furthermore, we find regimes in which the attack rate has local optima, minima or monotonously varies with the mixing rate. We corroborate our findings with Monte Carlo simulations on a primary-school network. This study provides a mathematical basis to better understand how homophily in health behavior shapes the dynamics of epidemics.
\end{abstract}

\maketitle


Preventing disease outbreaks is one of the greatest challenges humanity faced in recent history \cite{plagues}. One of the techniques employed to fight epidemics is identifying and eventually testing or isolating the contacts of infected individuals, which is generally referred to as contact tracing \cite{Keeling2011}. In the past, contact tracing was employed to combat the spread of smallpox, tuberculosis, HIV or Ebola \cite{foege1971selective, Fox2013, Rutherford1988,Swanson2018}, to name a few. Accordingly, great efforts have been devoted to analyze the efficacy of manual contact tracing  \cite{HOWELL1997,Kretzschmar1996, Fraser2004, Klinkenberg2006, Porco2004, Muller2000contact, Kiss2006, Eames2003, Eichner2003, Becker2005}. More recently, the advent of digital proximity tracing (DPT) apps, and their successful large-scale implementation to prevent the spread of SARS-COV-2 \cite{Rodriguez2021population, Braithwaite2020, Marcel2020, Fraser}, sparked numerous studies that analyzed how this novel technology impacts disease propagation. Many of these studies were tailored for SARS-COV-2 and quantified the impact of DPT apps \cite{Kucharski2020, Bradshaw2021, Aleta2020modelling, Ferretti2020, Barrat2020, Hellewell2020, lorch2020, Cencetti2020, pinotti2020}. However, also a variety of more theoretical studies unveiled the physics behind DPT \cite{PRRAdriana2021, PRRGiacomo2021, Kryven2021, kojaku2020}. 

A question which has not been addressed so far is how homofilic adoption between individuals affects the efficacy of DPT. Empirical studies suggests that the voluntary adoption of DPT apps strongly varies among the population. App adoption was shown to correlate with age, income and nationality \cite{pinotti2020, Wyl2020}. Accordingly, due to homophily, app adoption is much more probable for contacts of an app user (adopter) compared to a random individual. As a matter of fact, in Switzerland around $70 \%$ of the contacts among adopters were found to use the app as well, while average, national adoption is only $20 \%$ \cite{Marcel2020}. This discrepancy is in line with experimental studies, that indicate how homophily importantly affects health behavior \cite{Centola2010spread, Centola2011}. Such evidence let us theorize that this is another manifestation of how human behavior crucially shapes the course of epidemics \cite{Funk2010, Wang2016}. 

Our goal here is to follow this hypothesis and unveil how the homophilic adoption of DPT apps affects the disease propagation. For this aim, we extend a model recently introduced by Bianconi et al. \cite{PRRGiacomo2021}. The model is very convenient due to its simplicity, while it captures the essential ingredients of the dynamics. As a first step, we consider a mean field case, in which we control the interaction rate among adopters and non adopters. This simple setup shows different regimes, in which the homophilic adoption can be beneficial as well as detrimental. Furthermore, we analytically show how the reproduction number is minimized for a specific value of homophily. In a second step, we analyze the impact of homophily on a real world primary-school network. The network exhibits the same dynamical regimes and thus underpins the theoretical results. 
  
For modelling the epidemics we consider a standard SIR model, with transmission probability $\lambda$, infectious period $\tau$, and contact rate $k$. For convenience we define $\beta = \lambda \tau$. Accordingly, in the absence of app users, the basic reproduction number of the disease is given by $R_0 = \beta k$. We fix the fraction of adopters as $T\in\left[0,1\right]$. Furthermore, we parametrize the mixing relation between adopters and non adopters, i.e. the contact matrix $K$ (often referred to as the who-acquires-infection-from-whom matrix \cite{Keeling2011}), with a parameter $\alpha \in [0,1]$. We denote the entries of $K$ as $k_{ij}$ with  $i,j \in \{\text{A}, \text{N}\}$, where A and N refer to adopters and non adopters, respectively. The parameter $\alpha$ fixes the contact rate between adopters and non adopters as $k_{\text{AN}} = \alpha k  (1-T)$. In other words, $\alpha$ interpolates from complete homophily ($\alpha = 0$) to random mixing ($\alpha = 1$).

This parametrization does not allow for disassortativity. However, since the empirical evidence clearly indicates a positive correlation between social contacts and app adoption, we discard this possibility. Eventually, the remaining contact rates follow from the balance equation $T k_{\text{AN}} = (1-T)k_{\text{NA}}$ and the average contact rate $k = k_{\text{AA}} + k_{\text{AN}} = k_{\text{NN}} + k_{\text{NA}}$. Accordingly, $K$ has the following entries 
\vspace{0.5cm}
\begin{align}
    k_{\text{AN}} &= \alpha (1-T)k \\
    k_{\text{NA}} &= \alpha T k \\
    k_{\text{AA}} &= \left[1 - \alpha (1-T) \right]k \\
    k_{\text{NN}} &= \left[ 1 - \alpha T\right]k  \,.
\end{align}

\newpage

\onecolumngrid

\begin{figure}
    \includegraphics[width = 1.0\linewidth]{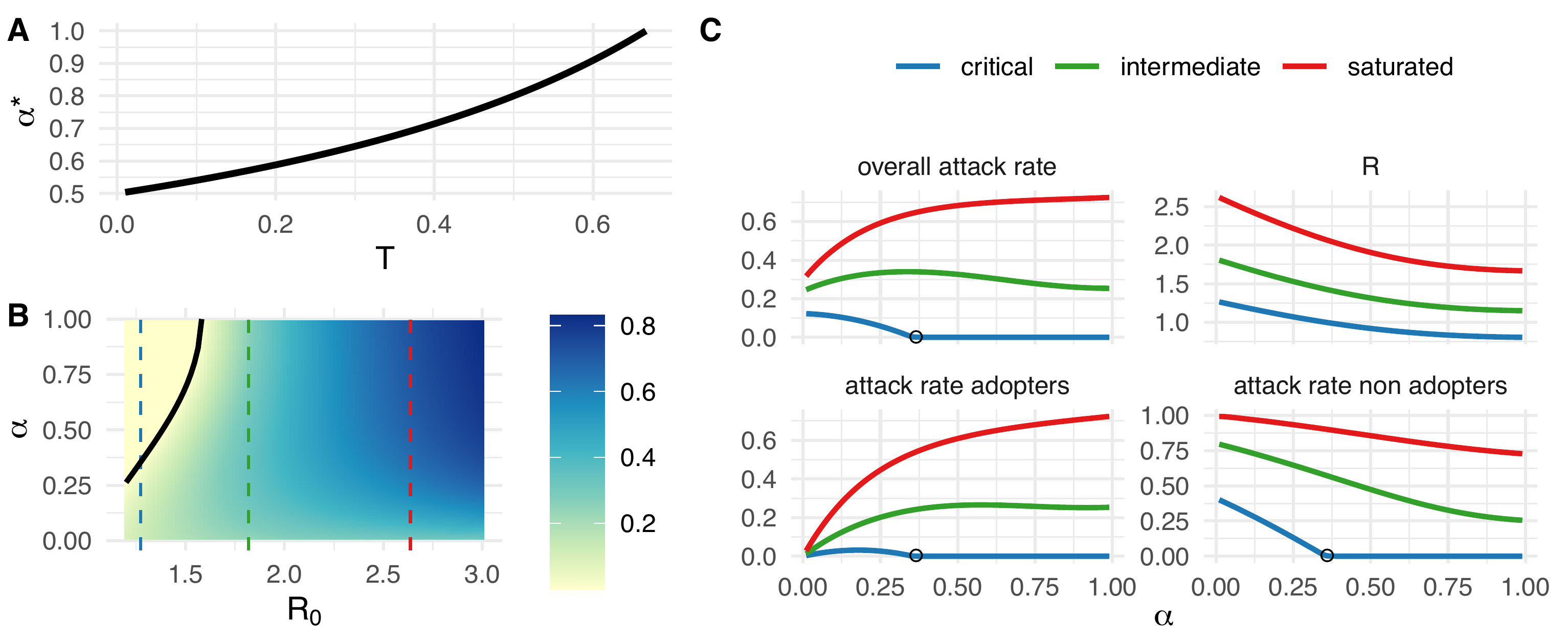}
    \caption{{\bf A:} Mixing rate that minimizes the reproduction number $\alpha^{*}$ as a function of adoption, $T$. {\bf B:} Attack rate as a function of $\alpha$ and the basic reproduction number of the disease, $R_0$. The solid line indicates the threshold $\alpha_c^{-}$ for which $R = 1$. Colored, dashed lines denote different dynamical regimes. Adoption was fixed as $T = 0.7$. For this parameter values we have $\alpha_c^{+} > 1$. {\bf C:} Top panels show the attack rate and the reproduction number along the alike colored dashed lines in B. The specific attack rates for adopters and non adopters are reported in the bottom panels. Black circles indicate $\alpha_c^{-}$.}
    \label{fig:heatmapRAB}
\end{figure}
\twocolumngrid

\noindent With these definitions, the probability that during a contact both individuals are either adopters or not reads $1 - 2\alpha T(1-T)$. Correctly, this probability becomes $1$ for complete isolation ($\alpha = 0$) and $1-2T(1-T)$ for random mixing ($\alpha = 1$).

The dynamical model introduced by Bianconi et. al \cite{PRRGiacomo2021} is based on the assumption that app users infected by other app users, do not further transmit the disease. This assumption requires us to define additional variables to follow the dynamical evolution in comparison to the standard SIR model. Namely, we define $I_{\text{AA}}(t)$ and $I_{\text{AN}}(t)$ as the number of newly infected (incidence) adopters at generation $t$, which were infected by adopters and non adopters, respectively. For non adopters such distinction is not necessary, wherefore $I_{\text{N}}(t)$ simply refers to the newly infected non adopters at generation $t$. Together with the number of susceptible adopters $S_{\text{A}}(t)$ and non adopters $S_{\text{N}}(t)$ at generation $t$, the discrete dynamical equations then read

\begin{align}
    I_{\text{N}}(t+1) &= \left[ k_{\text{NN}}  I_{\text{N}}(t) +   k_{\text{AN}}  I_{\text{AN}}(t)\right] \frac{S_{\text{N}}(t)}{N_{\text{N}}} \\
    I_{\text{AN}}(t+1) &=  \beta k_{\text{NA}}  I_{\text{N}}(t) \frac{S_{\text{A}}(t)}{N_{\text{A}}} \\
    I_{\text{AA}}(t+1) &=  \beta k_{\text{AA}}  I_{\text{AN}}(t) \frac{S_{\text{A}}(t)}{N_{\text{A}}} \\
    S_{\text{N}}(t+1) &= S_{\text{N}}(t) - I_{\text{N}}(t+1) \\
    S_{\text{A}}(t+1) &= S_{\text{A}}(t) - I_{\text{AA}}(t+1) - I_{\text{AN}}(t+1) \,. 
\end{align}

\noindent The parameters $N_{\text{A}}$ and $N_{\text{N}}$ indicate the number of adopters and non adopters, respectively, in the population. The attack rate, i.e. the number of recovered individuals, is found by summing the number of newly infected (then recovered) individuals, over all generations.

As the classical SIR model, this dynamical system does not allow for an explicit solution of the non trivial, stationary state. However, the model allows to calculate the reproduction number, $R$, for a fully susceptible population. For $R > 1$, the disease initially invades the population whereas for $R < 1$ it immediately dies out. The reproduction number can be calculated through the next-generation matrix \cite{Diekmann1990, Diekmann2009}, NGM, which is given by

\begin{equation}
    \text{NGM} = \begin{pmatrix}
    \beta k_{\text{NN}} & \beta k_{\text{AN}} \\
    \beta k_{\text{NA}} & 0 
\end{pmatrix} \,.
\end{equation}

\noindent The zero entry of the NGM is due to the tracing capacity preventing adopters to cause any new infections (generations), when infected by other adopters. The largest eigenvalue, i.e. the spectral radius of the NGM represents the effective reproduction number, $R$. By inserting the explicit expressions of the $K$ matrix entries, the effective reproduction number reads

\begin{equation}
\label{eq:repNumber}
    R = \frac{R_0}{2} \left[ 1-\alpha T + \sqrt{(1-\alpha T)^2 + 4\alpha^2 T (1-T)} \right] \,.
\end{equation}

\noindent Not surprisingly, $R$ has a monotonous dependence on adoption, $T$, as well as on the basic reproduction number of the disease, $R_0$. However, $R$ exhibits a non trivial dependence on the mixing rate, $\alpha$. To be more specific, solving $\frac{dR}{d\alpha} = 0$ with respect to $\alpha$ yields

\begin{equation}
    \alpha^{*} = 1 + \frac{T - \frac{2}{3}}{\frac{4}{3} - T} \,.
\end{equation}

\noindent Straightforward calculations also show that $\frac{dR}{d\alpha}|_{\alpha = 0} < 0$ is always met. Accordingly, whenever $\alpha^{*} \geq 1$, the reproduction number has its smallest value at $\alpha = 1$. The condition $\alpha^{*} < 1$ leads to a critical value $T_c = 2/3$ above which no local minimum exists. Fig.~\ref{fig:heatmapRAB}A shows how $\alpha^{*}$ varies with respect to adoption, $T$.

This non monotonous dependence on the mixing rate, $\alpha$, distinguishes contact tracing from classical immunization problems as vaccination \cite{Salathe2008}. Assuming perfect immunization, all the entries of the NGM will be zero, except the one among not immunized individuals (non adopters). Accordingly, the dependence on $\alpha$ will be monotonous and mixing will always reduce the reproduction number. 

Furthermore, Eq.~\eqref{eq:repNumber} provides a critical parameter range in which eradication is possible, i.e. $R < 1$. Since our focus is the effect of homophily, we express the condition for eradication as a function of $\alpha$. In other words for any $\alpha \in (\alpha_c^{-}, \alpha_c^{+})$, we have $R < 1$, where 

{\small
\begin{equation}
    \alpha_c^{\pm} = \frac{1}{2R_0 (1-T)}\left(1 \pm \sqrt{1 - 4 \frac{1-T}{T}(R_0 - 1)} \right) \,.
\end{equation}
}

\noindent The existence of two physical solutions of $\alpha_c$ implies that increasing mixing may not only push the system below threshold, but also push it above for $\alpha > \alpha_c^{+}$ and thus hinder eradication. This possibility emerges from the existence of a local minimum in the reproduction number with respect to the mixing rate, $\alpha$.

\begin{figure}[b]
    \centering
    \includegraphics[width = 0.8\linewidth]{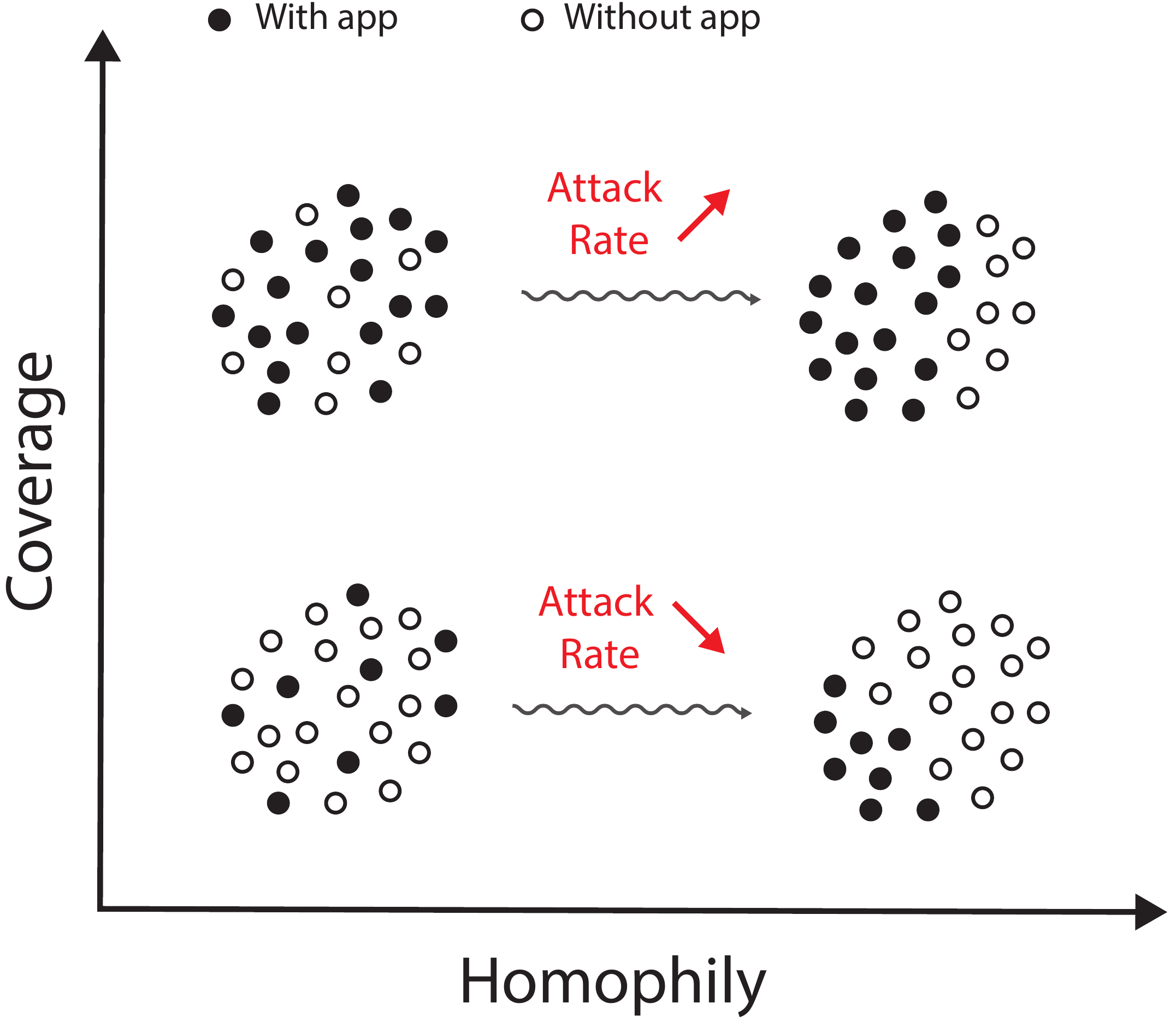}
    \caption{Schematic illustration of the interplay between homophily and coverage in our model.}
    \label{fig:dibujo}
\end{figure}

Fig.~\ref{fig:heatmapRAB}B shows the attack rate as a function of $\alpha$ and the basic reproduction number, $R_0$. The solid line indicates $\alpha_c^{-}$. The attack rate monotonously increases with $R_0$. More interestingly, as indicated by the dashed lines, $R_0$ separates three different regimes regarding the dependence of the attack rate on $\alpha$.

In Fig.~\ref{fig:heatmapRAB}C we explicit these three regimes by presenting the attack rate as a function of $\alpha$ following the dashed lines. We categorize these regimes, in ascending order with respect to $R_0$, as critical, intermediate and saturated. Close to threshold (critical), the attack rate decreases with the mixing between adopters and non adopters. In contrast, in the intermediate regime, we observe a non monotonous dependence of the attack rate on $\alpha$. Finally, far from the epidemic threshold, in a saturated regime, the attack rate continuously increases with $\alpha$.

\begin{figure}[t]
    \centering
    \includegraphics[width = 0.9\linewidth]{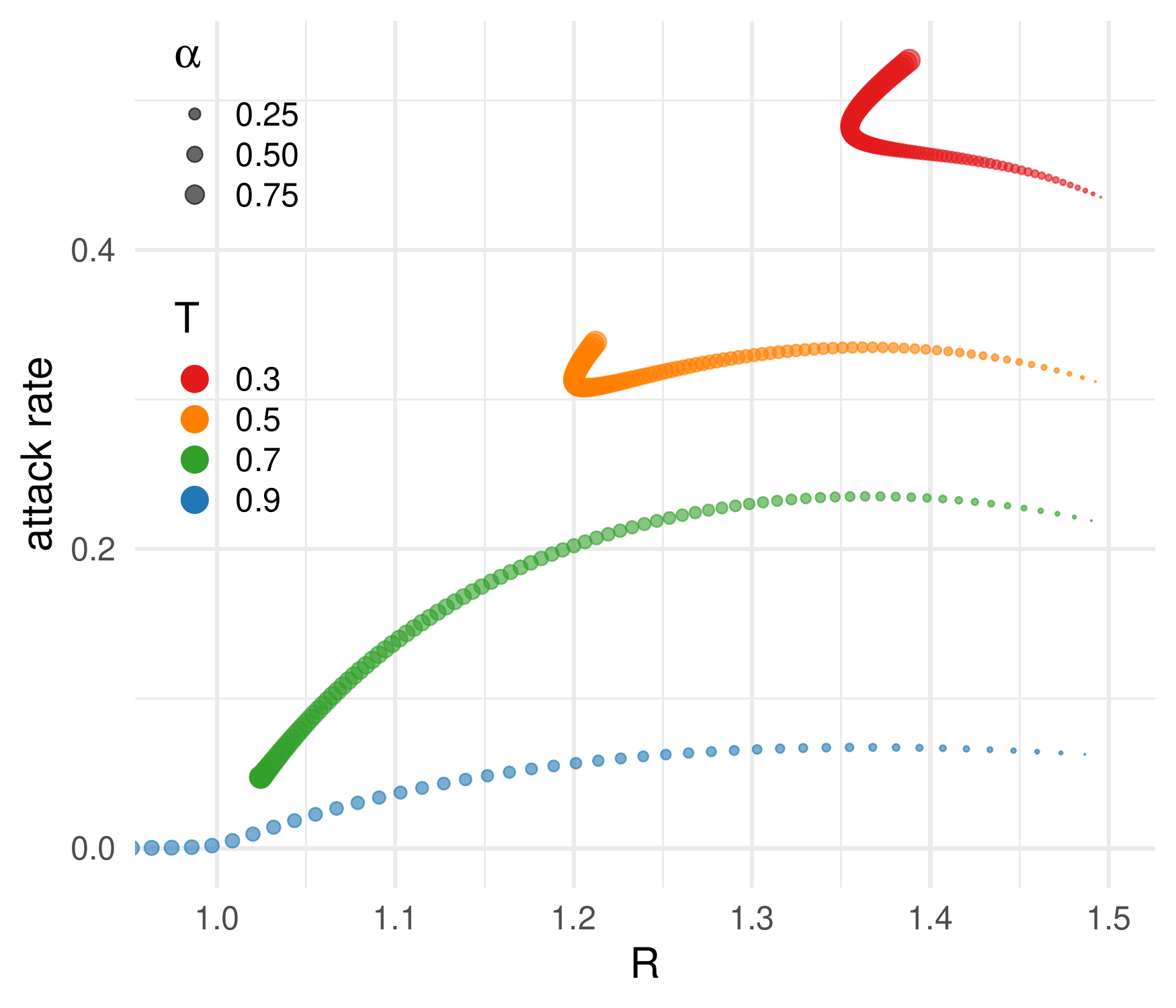}
    \caption{Attack rate as a function of the reproduction number, $R$, for different values of adoption, $T$. Size of the points interpolates between $\alpha = 0$ and $\alpha = 1$. We fixed the basic reproduction number as $R_0 = 1.5$}
    \label{fig:nontrivial}
\end{figure}

The variety of regimes stems from the competition between two processes. On the one hand, as mixing increases, adopters provide protection to non adopters. This is illustrated by a decreasing reproduction number, $R$, (top-right panel in Fig.~\ref{fig:heatmapRAB}C) and attack rate among non adopters (bottom-right panel in Fig.~\ref{fig:heatmapRAB}C). On the other hand, protection vanishes inside the adoption cluster (bottom-left panel in Fig.~\ref{fig:heatmapRAB}C). Adoption and infectivity, i.e. the basic reproduction number, $R_0$, then determine which of these two processes holds the upper hand. The schema displayed in Fig.~\ref{fig:dibujo} illustrates both processes. For low coverage, complete, homophilic adoption is more beneficial, since random mixing does not provide any protection -- neither to non adopters nor to adopters. In contrast, for high coverage, random distribution of adopters acts as a firewall for non adopters and enables to immunize the population.

Whether a given adoption is sufficient to provide protection to non adopters is determined by the basic reproduction number, $R_0$. Far from the critical threshold, in the saturated regime, the attack rate among non adopters only slowly varies with the reproduction number, $R$. Accordingly, mixing increases the attack rate. In contrast, close to the threshold, in the critical regime, the attack rate among non adopters strongly varies with a decrease in the reproduction number. Therefore, mixing is beneficial and allows to push the system below the critical point, from which then also adopters benefit. Finally, in the intermediate regime, the system switches between critical and saturated, wherefore we observe a non monotonous dependence with respect to $\alpha$. 
\begin{figure}[t]
    \includegraphics[width = 1.0\linewidth]{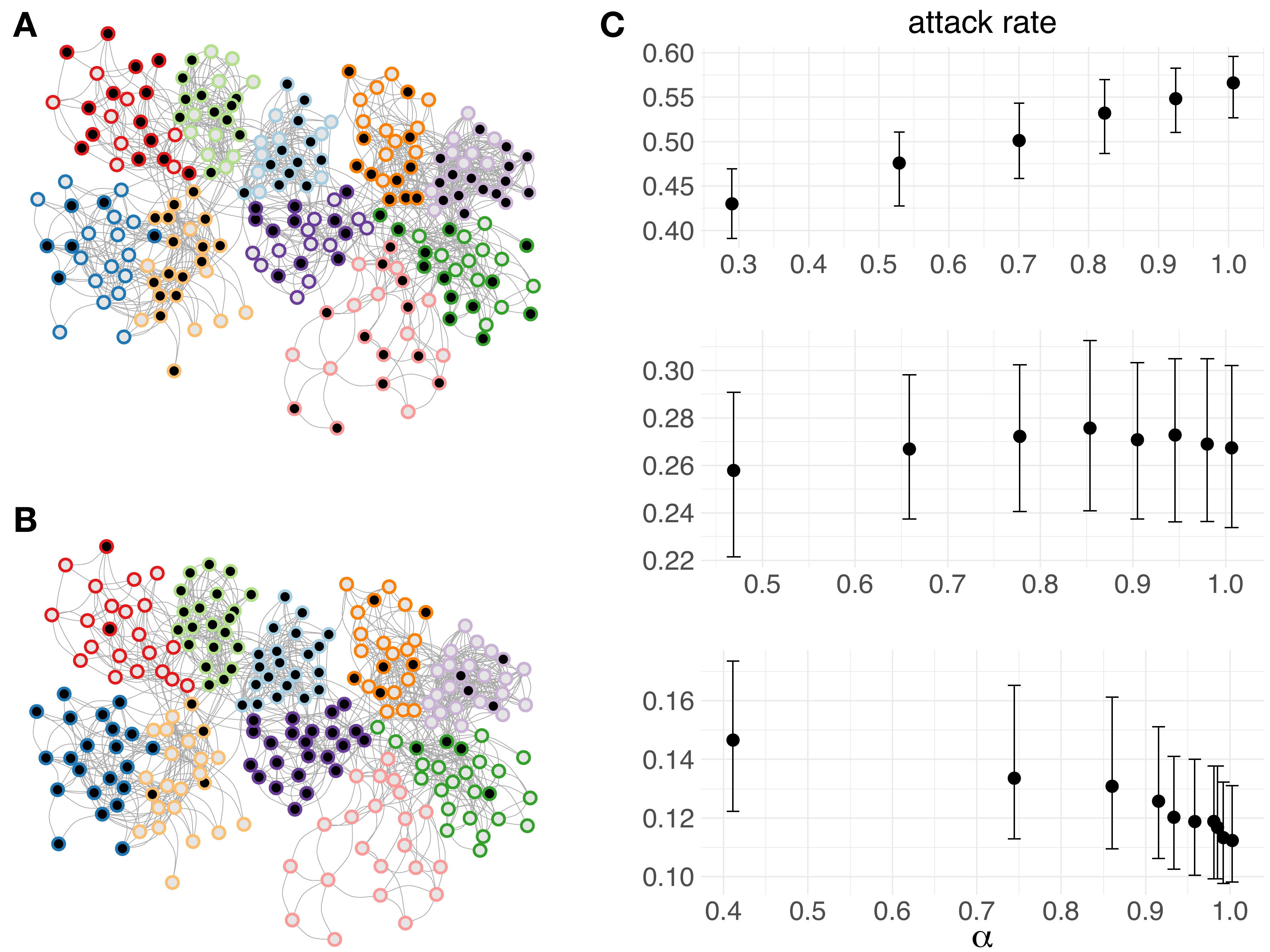}
    \caption{{\bf A:} Real-world primary-school contact network. Color borders refer to the community a node is member of, while its internal color indicates whether it adopted (black) or not (light grey) the tracing app. Here $T=0.5$, $m=0$ (random adoption). {\bf B:} Same as A but for $m=4$ (homophilic adoption). {\bf C:} Dependence of the attack rate on $\alpha$ as resulting from Monte Carlo simulations; $\beta=0.2$. Dots indicate the median value, whereas error bars indicate the first and third quartiles. Each point is obtained by averaging over $2.5 \times 10^4$ runs. The app coverage is fixed to $T = 0.52$ (top), $T = 0.77$ (middle) and $T = 0.92$ (bottom). We fixed $\beta = 0.2$. With an average degree of $11$, this then results in $R_0 = 2.2$.}
    \label{fig:highschool}
\end{figure}

We kept the coverage of the app, $T$, fixed in the above analysis of the different regimes. To investigate the effect of coverage, we fix the basic reproduction number, $R_0$, and vary $\alpha$ for different values of $T$. Fig.~\ref{fig:nontrivial} shows the attack rate as a function of the reproduction number, $R$. While colors indicate different values of $T$, the mixing rate, $\alpha$, is mapped to point size. For high values of $T$ ($0.9$ \& $0.7$), the system is in between the intermediate and critical regime. Accordingly, we find a local maximum, but the disease can also be eradicated, $R < 1$. For lower values of $T$ ($0.5$ \& $0.3$), we observe a local minimum of the reproduction number since $T < T_c$. This local minimum in the reproduction number, then causes an additional minimum in the attack rate for $T = 0.5$. In contrast, for $T= 0.3$ the attack rate continuously increases with $\alpha$. More interestingly, due to the non monotonous form of $R$, we observe different attack rates for equal values of the reproduction number. Overall, Fig.~\ref{fig:nontrivial} illustrated how the three dynamical regimes can also be reached by varying adoption, $T$. 

To corroborate our theory, we used a real-world primary-school network \cite{network}. This physical contact network consists of $226$ nodes (students) grouped in $10$ communities (classes), whose edges are weighted by the time duration of proximity contacts. To transform the temporal network into a static, binary one, we threshold the aggregated weights. Normalizing with respect to the maximal weight, the relative threshold of $0.021$ was fixed just before the network would not have been connected anymore. The resulting network is shown in Fig.~\ref{fig:highschool}A \& \ref{fig:highschool}B for two dissimilar app coverings. To systematically study the effect of homophilic vs. random adoption, we leveraged the modular structure of the network. For a given value of app adoption, $T$, we entirely covered $m\leq \lfloor 10T \rfloor$ communities with app, while randomly distributing the remaining fraction of apps, $T-m/10$, through all the other communities. For each value of app coverage, $T$, we computed the adoption homophily $h$ as the fraction of edges whose end nodes where both app or non app users. The mixing parameter $\alpha$ is then obtained as

\begin{equation}
    \alpha = \frac{1-h}{2T(1-T)} \,.
\end{equation}

\noindent This correlates negatively with $m$, so that $m=0$ corresponds to random adoption, $\alpha \approx 1$, while $m= \lfloor 10T \rfloor$ to complete homophilic adoption, hence low values of $\alpha$. Fig.~\ref{fig:highschool}C reports the results of Monte Carlo simulations for the attack rate versus $\alpha$. The curves, each corresponding to a different value of $T$, show the non trivial phenomenological change through the three dynamical regimes of the epidemics, i.e. saturated, intermediate and critical, as correctly identified by our analytical model.

To sum up, we analyzed how homophilic adoption of DPT apps affect the disease dynamics. We unveiled the existence of different dynamical regimes, originating from a non trivial dependence on the mixing rate. In the critical regime, mixing is beneficial, may enable to push the system below threshold and thus eradicate the disease. Far from the threshold, mixing is detrimental due to a waning protection among app users. Finally, for an intermediate case, the system switches between the two regimes with varying mixing rate, and the dependence is non monotonous. Moreover, we discovered a local minimum in the reproduction number, existing whenever adoption is smaller than $2/3$. Accordingly, an increasing mixing rate may even push the system above the critical threshold and cause the disease to spread. Interestingly, the different regimes in the attack rate can arise independently on whether the reproduction has a local minimum or not.

Switching our focus to the real world, adoption of DPT apps is generally very low, between $20 \%$ and $40 \%$ \cite{Marcel2020, Fraser}. Accordingly, it is very improbable to control the epidemic with such low adoption. In this sense, our results indicate that homophilic adoption is beneficial to this point. However, if health authorities desire to actually contain the spread of SARS-COV-2, i.e. reach the critical regime, overcoming homophily in health behavior to have a more homogeneous distribution may prove crucial. Overall, this study should contribute to the mathematical foundation on how heterogeneity in human behavior affects the course of epidemics and spark new theoretical studies. A preliminary analysis indicates that equivalent dynamical regimes exists in the case of imperfect immunization, i.e. vaccination, but also for any other prophylactic measures, such as social distancing or the use of face masks. \\

\emph{Acknowledgments.} G.B. acknowledges financial support from the European Union’s Horizon 2020 research and innovation program under the Marie Sk\l{}odowska-Curie Grant Agreement No. 945413 and from the Universitat Rovira i Virgili (URV). B.S. acknowledges financial support from the European Union’s Horizon 2020 research and innovation program under the Marie Sk\l{}odowska-Curie Grant Agreement No. 713679 and from the Universitat Rovira i Virgili (URV). A.A. acknowledges support by Ministerio de Econom\'ia y Competitividad (Grants No. PGC2018- 094754-B-C21 and No. FIS2015-71582-C2-1), Generalitat de Catalunya (Grant No. 2017SGR-896), Universitat Rovira i Virgili (Grant No. 2019PFR-URV-B2-41), Spanish MINECO (grant PGC2018-094754-B-C21), ICREA Academia, and the James S. McDonnell Foundation (Grant No. 220020325). We thank L. Arola-Fernández for helpful comments and suggestions.

G.B. and B.S. contributed equally to this work.

\bibliography{apssamp}

\end{document}